\RequirePackage{fix-cm}
\RequirePackage{amsmath}

\documentclass[12pt,prd,nofootinbib]{revtex4-1}
\usepackage[utf8]{inputenc} 

\usepackage{CJK}
\usepackage[T1]{fontenc}
\usepackage[left=2.5cm,top=3.0cm,right=2.5cm,bottom=3.0cm]{geometry} 
\usepackage{array}
\usepackage{textcomp}
\usepackage{multirow}
\usepackage{amsmath}
\usepackage{graphicx}
\usepackage{hhline}
\usepackage{breqn}
\usepackage{comment}
\usepackage{adjustbox}
\usepackage{support-caption}
\usepackage{subcaption}
\usepackage{amsmath}
\usepackage{amsfonts}
\usepackage{amssymb}
\usepackage{cancel} 
\usepackage{youngtab}
\usepackage{xfrac}

\usepackage{pstricks}
\usepackage{url}
\usepackage{color}

\usepackage{graphicx}
\usepackage{feynmp} 
\usepackage{float} 

\begin{document}

\title{PMNS matrix in a non-universal $U(1)_{X}$ extension to the MSSM with one massless neutrino}




\author{J.S. Alvarado,} \thanks{jsalvaradog@unal.edu.co}
\author{R. Martinez. } \thanks{remartinezm@unal.edu.co}

\affiliation{Departamento de Física$,$ Universidad Nacional de Colombia\\
 Ciudad Universitaria$,$ K. 45 No. 26-85$,$ Bogot\'a D.C.$,$ Colombia}

\date{\today}

\begin{abstract}
An anomaly free non-universal $U(1)_{X}$ extension to the Minimal Supersymmetric Standard Model is proposed, where additional two $SU(2)$ doublet superfields and four singlet superfields complement the scalar sector of the model. The fermion sector is extended by considering additional superfields containing three quark singlets, two charged lepton singlet superfields and six neutral leptons. A tree-level massless electron is found so radiative corrections are considered to match the mass spectrum with both SUSY and non-SUSY contributions. Likewise, a massless neutrino is found and analytic expressions for massive mass eigenstates are obtained via inverse-seesaw mechanism, which implies a known neutrino mass spectrum for both normal and inverse ordering. Lastly, a numerical fitting of the model parameters to the PMNS matrix is done.

\textbf{Keywords:} Extended scalar sectors, Supersymmetry, Beyond the standard model, Fermion masses, PMNS matrix, Exotic fermions. 
\end{abstract}
\maketitle 

\section{Introduction}\hfill

It is known that the Standard Model (SM) agrees with almost all the experimental results although there is no evidence of new physics yet. For instance, electroweak spontaneous symmetry breaking (SSB) accounts for fermion masses in the current SM due to a single vacuum expectation value (VEV) at the GeV scale. However, the $\sim 10^{6}$ difference between electron and top quark masses lead to the possibility to consider new physics beyond the SM that can provide an explanation of such masses with Yukawa couplings of the same order. This issue, called fermion mass hierarchy (FMH), has been a motivation to extend the SM by adding new particles or symmetries. Moreover, neutrino masses are a currently unexplained problem in the SM due to the absence of right-handed neutrinos, but neutrino masses are experimentally confirmed by neutrino oscillation experiments. The latter problems have encouraged several SM extensions in an attempt of predicting physics beyond the SM.

For instance, supersymmetry (SUSY) is a promising extension which has theoretical attraction for Grand Unification Theories (GUT) \cite{SUSYGUT}, string theories and Dark Matter candidates \cite{SUSYDM}. It also provides a scenario for a natural realization of FMH, Leptonic Flavor Violation (LFV)\cite{ellis} and Flavor Changing Neutral Currents (FCNC)\cite{FCNC} in non-universal extensions. Additionally, SUSY solves the Higgs naturalness problem, but it has its own inconveniences such as the $\mu$ problem \cite{muproblem}. Nevertheless, it is still relevant to study some unexplored features of SUSY theories \cite{SUSY}.

The most known SM result is the discovery of the Higgs boson \cite{Higgs} which together with the $W$ and $Z$ gauge boson masses confirm the underlying  $SU(3)_{C} \times SU(2)_{L}\times U(1)_{Y}$ symmetry before spontaneous symmetry breaking \cite{weinberg}-\cite{weak}. Despite different Yukawa couplings can explain fermion masses and they are stable under radiative corrections, there is always the possibility of having a more natural scenario for such a huge difference among fermion masses. There are non-universal extensions in both SUSY and non-SUSY scenarios taking into account multiple Higgses models,  such as the Two Higgs doublet model (2HDM)\cite{2HDM} and new scalar fields \cite{vevstable1}-\cite{vevstable3}.

Non-universality requires an extra symmetry if fine-tuning is avoided. By climbing a step we can propose a $U(1)_{X}$ extension to the SM so this new $X$ charge can have a non-universal character. These kinds of models have been widely studied \cite{abelian} and they usually include an approach for FMH and neutrino mass generation. The main non-universality feature lies in the existence of FCNC whose experimental sensitivity leaves an open window for new physics, such as the $B$ meson decay \cite{FCNClep} and Lepton Flavor Violation \cite{LFVnew}. 

Nevertheless, neutrino physics is one of the most promising scenarios for physics beyond the standard model since they are massive particles, as neutrino oscillation has proven, and its nature is to date unknown. In fact, the already known particles are unable of providing a scenario for neutrino mass generation so new heavy particles are considered in seesaw mechanisms \cite{seesaw},\cite{inverseseesaw} as an alternative for the neutrino mass generation. Even so, non-universality introduces different mass matrix textures which not always guarantees a correct PMNS reproducibility \cite{neutrinoTex}, \cite{MTex}.

The present work is derived from 331 models \cite{331model} where the $SU(3)_{L}\times U(1)_{X}$ symmetry, for $\beta=-\sfrac{1}{\sqrt{3}}$, is replaced by a simpler gauge structure as $SU(2)_{L} \times U(1)_{Y} \times U(1)_{X}$ but preserving the $X$ charges. In this way, we can keep the same fermionic content and anomaly cancellation. We introduce the fermionic sector in the non-universal $U(1)_{X}$ extension to the Minimal Supersymmetric Standard Model (MSSM) whose extended superfield content leads to a scalar sector containing four doublets and four singlets while the fermion sector contains an additional up-like quark singlet, two down-like quark singlets, two charged lepton singelts and three generations of right-handed and Majorana neutrinos\footnote{We consider squarks and sleptons as not contained in the scalar sector. Likewise, gauginos and Higgsinos are not included in the fermion sector}. The conditions and analytic expressions for mass eigenstates and rotation matrices are obtained, letting us to check the reproducibility of the PMNS matrix, as well as Yukawa couplings predictions for a wide range of values.

Lepton-neutrino mixing happens in a similar fashion as quark mixing.  In such case, the relative rotation between lepton and neutrino mass eigenstates is known as the Pontecorvo–Maki–Nakagawa–Sakata \cite{PMNSmatrix} (PMNS) matrix which is given analogously to the CKM matrix by $V_{PMNS}=V_{l}U_{\nu}^{\dagger}$. 

Despite the CKM and PMNS matrices are mathematically analogous, the neutrino mass matrix entries are much smaller than the quark matrix entries since experimentally it is known that neutrinos must have a very small mass \cite{noscillations}\cite{noscillationstheory}. Consequently, the oscillations between quarks in the flavor basis happens in small times, corresponding to a length scale of the nuclear radius while neutrino oscillations require hundreds of kilometers  as pointed out in the solar neutrino problem \cite{solarnu} resulting in highly suppressed flavor changes ($l_{\alpha}\rightarrow l_{\beta} < 10^{-54}$) \cite{Lsupression}. Currently, the PMNS matrix has the following magnitude values \cite{nufit} at $3\sigma$ confidence level: 
\begin{align}
    V_{PMNS}&=\begin{pmatrix}
0.801 \rightarrow 0.845 & 0.513 \rightarrow 0.579 & 0.143 \rightarrow 0.156\\
0.233 \rightarrow 0.507 & 0.461 \rightarrow 0.694 & 0.631 \rightarrow 0.778\\
0.261 \rightarrow 0.526 & 0.471 \rightarrow 0.701 & 0.611 \rightarrow 0.761\\
\end{pmatrix}.\label{PMNSexp}
\end{align}

\begin{table}[H]
    \centering
    \begin{tabular}{|c|c|c|} \hline \hline
         & NO & IO \\ \hline
        $\frac{\Delta m_{21}^{2}}{10^{-5}eV^{2}}$ & $7.42_{-0.20}^{+0.21}$ & $7.42_{-0.20}^{+0.21}$ \\ \hline
        $\frac{\Delta m_{3\ell}^{2}}{10^{-3}eV^{2}}$ & $+2.514_{-0.027}^{+0.028}$ & $-2.497_{-0.028}^{+0.028}$ \\ \hline 
        $\theta_{12} /^{\circ}$ & $33.44_{-0.75}^{+0.78}$  &  $33.45_{-0.75}^{+0.78}$  \\ \hline
        $\theta_{23}/^{\circ}$ & $49.0_{-1.4}^{+1.1}$ & $49.3_{-1.2}^{+1.0}$ \\ \hline
        $\theta_{13}/^{\circ}$ & $8.57_{-0.12}^{+0.13}$ &  $8.61_{-0.12}^{+0.12}$ \\ \hline
        $\delta /^{\circ}$ & $195_{-25}^{+51}$ & $286_{-32}^{+27}$ \\ \hline
    \end{tabular}
    \caption{Neutrino mixing parameters and squared mass differences for normal and inverse ordering \cite{nufit}.}
    \label{tablenu}
\end{table}

The elements of the PMNS matrix are determined mainly from neutrino oscillation experiments by studying the three main sources: solar, atmospheric and reactor neutrinos. They get the mixing angles and the CP phase, given in table \ref{tablenu}. Nevertheless, they only provide squared mass differences instead of information about each mass eigenstate. In general, two schemes are considered: the normal ordering (NO) ( $m_{1} \lesssim m_{2}< m_{3}$) and the inverse ordering (IO) ($m_{3} < m_{1} \lesssim m_{2}$). It is important to mention the mixing matrices standard parametrization in terms of three mixing angles and one CP phase, in the following form:
\begin{equation}\label{standardparam}
    U=\begin{pmatrix}
    1 & 0 & 0 \\
     0 & \cos\theta_{23} & \sin\theta_{23} \\
     0 & -\sin\theta_{23} & \cos\theta_{23}
    \end{pmatrix}
    \begin{pmatrix}
    \cos\theta_{13} & 0 & \sin\theta_{13}e^{-i\delta} \\
    0 & 1 & 0 \\
    -\sin\theta_{13}e^{i\delta} & 0 & \cos\theta_{13}
    \end{pmatrix} 
    \begin{pmatrix}
    \cos\theta_{12} & \sin\theta_{12} & 0 \\
    -\sin\theta_{12} & \cos\theta_{12} & 0 \\
    0 & 0 & 1 
    \end{pmatrix}.
  \end{equation}

\section{The $U(1)_{X}$ extension}

Most of the models predict a new gauge boson called $Z'$ where this particle exist from a $U(1)$ additional symmetry \cite{U1}, as part of a non-abelian higher symmetry  such as $SU(2)_{R}$, $331$ models \cite{nonabelian}, Kaluza-Klein excitations in extra dimensions \cite{KK} or as a string resonance \cite{string}. In the first two scenarios, a higher symmetry group that breaks into an additional $U(1)$ gauge symmetry can provide an explanation to fermion mass hierarchy.

A new non-universal $U(1)_{X}$ interaction together with a $\mathcal{Z}_{2}$ parity are included into the MSSM based on the non-supersymmetric version of the model \cite{nosusy} which provides a scenario for understanding FMH based on the existence of two Higgs doublets and two scalar singlets. In the fermionic sector, there is an exotic up-like quark ($\mathcal{T}$), two down-like quarks ($\mathcal{J}^{a}$, $a=1,2$), two exotic leptons ($E$, $\mathcal{E}$), three right-handed neutrinos ($\nu_{L}^{C}$) and three heavy Majorana neutrinos ($N_{R}$) all of them interacting via the scalar singlets and generating a mass matrix texture compatible with FMH.

When a new symmetry is included in a theory, there is always the risk of inducing anomalies so the inclusion of a new $U(1)$ symmetry leads to the following set of equations for the X charges in such a way that the theory remains anomaly free:
\begin{eqnarray}
\left[\mathrm{\mathrm{SU}(3)}_{C} \right]^{2} \mathrm{\mathrm{U}(1)}_{X} \rightarrow & A_{C} &= \sum_{Q}X_{Q_{L}} + \sum_{Q}X_{Q_{L}^{c}},	\label{an1}	\\
\left[\mathrm{\mathrm{SU}(2)}_{L} \right]^{2} \mathrm{\mathrm{U}(1)}_{X} \rightarrow & A_{L}  &= \sum_{\ell}X_{\ell_{L}} + 3\sum_{Q}X_{Q_{L}}	,	\\
\left[\mathrm{\mathrm{U}(1)}_{Y} \right]^{2}   \mathrm{\mathrm{U}(1)}_{X} \rightarrow & A_{Y^{2}}&=
	\sum_{\ell, Q}\left[Y_{\ell_{L}}^{2}X_{\ell_{L}}+3Y_{Q_{L}}^{2}X_{Q_{L}} \right]	
	+ \sum_{\ell,Q}\left[Y_{\ell_{L}^{c}}^{2}X_{L_{L}^{c}}+3Y_{Q_{L}^{c}}^{2}X_{Q_{L}^{c}} \right],		\\
\mathrm{\mathrm{U}(1)}_{Y}   \left[\mathrm{\mathrm{U}(1)}_{X} \right]^{2} \rightarrow & A_{Y}&=
	\sum_{\ell, Q}\left[Y_{\ell_{L}}X_{\ell_{L}}^{2}+3Y_{Q_{L}}X_{Q_{L}}^{2} \right]	
	+ \sum_{\ell, Q}\left[Y_{\ell_{L}^{c}}X_{\ell_{L}^{c}}^{2}+3Y_{Q_{L}^{c}}X_{Q_{L}^{c}}^{2} \right],		\\
 \left[\mathrm{\mathrm{U}(1)}_{X} \right]^{3} \rightarrow & A_{X}&=
	\sum_{\ell, Q}\left[X_{\ell_{L}}^{3}+3X_{Q_{L}}^{3} \right]	
	+ \sum_{\ell, Q}\left[X_{\ell_{L}^{c}}^{3}+3X_{Q_{L}^{c}}^{3} \right] 	,	\\	
\left[\mathrm{Grav} \right]^{2}   \mathrm{\mathrm{U}(1)}_{X} \rightarrow & A_{\mathrm{G}}&=
	\sum_{\ell, Q}\left[X_{\ell_{L}}+3X_{Q_{L}} \right]
	+ \sum_{\ell, Q}\left[X_{\ell_{L}^{c}}+3X_{Q_{L}^{c}} \right] \label{an6}.
\end{eqnarray}

The 331 models are anomaly free when three families are considered. One possibility is to have two quark families in the $\textbf{3}$ representation of $SU(3)_{L}$, being 6 multiplets due to color multiplets, one quark family and three lepton families in the $\Bar{\textbf{3}}$ representation of $SU(3)_{L}$, resulting in 6 anti-multiplets as well, which implies that the particle content of $SU(3)_{C}\times SU(3)_{L}$ is vector like. The present abelian extension comes from the 331 model for $\beta=-\sfrac{1}{\sqrt{3}}$ where the fermionic content and $X$-charges are the same. Therefore, the model remains chiral anomaly-free. Nevertheless, anomaly equations were double-checked with Mathematica by considering the $X$-charge assignation shown in tables \ref{modelbosons} and \ref{modelfermions}. The main advantage of the $U(1)_{X}$ extension lies in the increased freedom to build mass matrix textures by changing fermion triplets by doublets and singlets and similarly for the scalar sector. Besides, all exotic gauge fields connecting SM and exotic fermions disappear, except by the extra neutral current. Nevertheless, the electromagnetic charge is defined in this model through the usual Gell-Mann-Nishijima relationship $Q=\mathcal{I}_{3}-\frac{1}{2}Y$ and it is worth to notice that due to the non-holomorphic interactions in SUSY, right-handed fields are represented by left-conjugate ones ($\bar{\psi_{R}}\rightarrow \psi_{L}^{c}$) making that right-handed particles in the model present the opposite electric charge. 

Although in the non-supersymmetric model \cite{nosusy} these equations are satisfied, when supersymmetry is imposed, they are not satisfied due to the presence of Higgsinos in the fermion content. The simplest way of avoiding this problem is by doubling the Higgs boson superfield content, so the additional superfields behave as the conjugate ones. The final particle content of the model is shown in tables \ref{modelbosons} and \ref{modelfermions}.

\begin{table}
\caption{Higgs boson superfield content of the model, hypercharge $Y$, non-universal $X$ quantum number, $\mathbb{Z}_{2}$ written in the form $X^{\mathbb{Z}_{2}}$.}
\label{modelbosons}
\centering
\begin{tabular}{lll cll}\hline\hline 
\multirow{3}{*}{
\begin{tabular}{l}
    Higgs    \\
    Scalar  \\
    Doublets
\end{tabular}
}
&\multicolumn{2}{l}{}&
\multirow{3}{*}{
\begin{tabular}{l}
    Higgs    \\
    Scalar  \\
    Singlets
\end{tabular}
} 
&\multicolumn{2}{l}{}\\ 
 &&&
 && \\ 
 &$X^{\pm}$&$Y$&
 &$X^{\pm}$&$Y$
\\ \hline\hline 
$\small{\hat{\Phi}_{1}=\begin{pmatrix}\hat{\phi}_{1}^{+}\\\frac{\hat{h}_{1}+v_{1}+i\hat{\eta}_{1}}{\sqrt{2}}\end{pmatrix}}$&$\sfrac{+2}{3}^{+}$&$+1$&
$\hat{\chi}=\frac{\hat{\xi}_{\chi}+v_{\chi}+i\hat{\zeta}_{\chi}}{\sqrt{2}}$	&	$\sfrac{-1}{3}^{+}$	&	$0$	\\
$\small{\hat{\Phi}_{2}=\begin{pmatrix}\hat{\phi}_{2}^{+}\\\frac{\hat{h}_{2}+v_{2}+i\hat{\eta}_{2}}{\sqrt{2}}\end{pmatrix}}$&$\sfrac{+1}{3}^{-}$&$+1$& $\sigma=\frac{\hat{\xi}_{\sigma}+i\hat{\zeta}_{\sigma}}{\sqrt{2}} $ &$ \sfrac{-1}{3}^{-} $ & $ 0 $		\\
$\small{\hat{\Phi}^\prime_{1}=\begin{pmatrix}\frac{\hat{h}_{1}'+v_{1}'+i\hat{\eta}_{1}'}{\sqrt{2}}\\\hat{\phi}_{1}^{-\prime}\end{pmatrix}}$&$\sfrac{-2}{3}^{+}$&$-1$&
$\hat{\chi}'=\frac{\hat{\xi}'_{\chi}+v_{\chi}'+i\hat{\zeta}'_{\chi}}{\sqrt{2}}$	&	$\sfrac{+1}{3}^{+}$ &	0\\
$\small{\hat{\Phi}^\prime_{2}=\begin{pmatrix}\frac{\hat{h}_{2}'+v_{2}'+i\hat{\eta}_{2}'}{\sqrt{2}}\\\hat{\phi}_{2}^{-\prime}\end{pmatrix}}$&$\sfrac{-1}{3}^{-}$&$-1$& $\sigma^{\prime} = \frac{\hat{\xi}_{\sigma}^{\prime}+i\hat{\zeta}_{\sigma}^{\prime}}{\sqrt{2}}$ & $\sfrac{+1}{3}^{-}$ &	0	\\\hline\hline
\end{tabular}
\end{table}

\begin{table}
\caption{Quark and lepton superfield content of the non-universal extension, hypercharge $Y$, $X$ quantum number and parity $\mathbb{Z}_{2}$ written in the form $X^{\mathbb{Z}_{2}}$.}
\label{modelfermions}
\centering
\begin{tabular}{lll lll}\hline\hline 
 Left-Handed Fermions &$X^{\pm}$&&
 Right-Handed Fermions &$X^{\pm}$&
\\ \hline\hline 
\multicolumn{6}{c}{SM Quarks }\\ 
\multicolumn{6}{c}{$Y_{q_{L}}=\sfrac{+1}{3}$, $Y_{u_{L}^{c}}=\sfrac{-4}{3}$, $Y_{d_{L}^{c}}=\sfrac{+2}{3}$}\\ \hline\hline
\begin{tabular}{c}	
	$ \hat{q} ^{1}_{L}=\begin{pmatrix}\hat{u}^{1}	\\ \hat{d}^{1} \end{pmatrix}_{L}$  \\
	$  \hat{q} ^{2}_{L}=\begin{pmatrix}\hat{u}^{2}	\\ \hat{d}^{2} \end{pmatrix}_{L}$  \\
	$  \hat{q} ^{3}_{L}=\begin{pmatrix}\hat{u}^{3}	\\ \hat{d}^{3} \end{pmatrix}_{L}$ 
\end{tabular} &
\begin{tabular}{c}
		$\sfrac{+1}{3}^{+}$	\\
	\\	$0^{-}$	\\
	\\	$0^{+}$	\\
\end{tabular}   &
\begin{tabular}{c}
			\\
	\\		\\
	\\		\\
\end{tabular}   &
\begin{tabular}{c}
	$ \begin{matrix}\hat{u}^{1\; c }_{L}	\\ \hat{u}^{2\; c}_{L} \end{matrix}$  \\
	$ \begin{matrix}\hat{u}^{3\; c}_{L}	\\ \hat{d}^{1\; c }_{L} \end{matrix}$  \\
	$ \begin{matrix}\hat{d}^{2\; c }_{L}	\\ \hat{d}^{3\; c }_{L} \end{matrix}$ 
\end{tabular} &
\begin{tabular}{c}
	$ \begin{matrix} \sfrac{-2}{3}^{+}	\\ \sfrac{-2}{3}^{-} \end{matrix}$  \\
	$ \begin{matrix} \sfrac{-2}{3}^{+}	\\ \sfrac{+1}{3}^{-} \end{matrix}$  \\
	$ \begin{matrix} \sfrac{+1}{3}^{-}	\\ \sfrac{+1}{3}^{-} \end{matrix}$ 
\end{tabular} &
\begin{tabular}{c}
	\\
	 \\
\end{tabular}
\\ \hline\hline 

\multicolumn{6}{c}{SM Leptons}\\ 
\multicolumn{6}{c}{$Y_{\ell_{L}}=-1$, $Y_{e_{L}^{c}}=+2$, $Y_{\nu_{L}^{c}}=0$}\\ \hline\hline	
\begin{tabular}{c}	
	$\hat{\ell}^{e}_{L}=\begin{pmatrix}\hat{\nu}^{e}\\ \hat{e} \end{pmatrix}_{L}$  \\
	$\hat{\ell}^{\mu}_{L}=\begin{pmatrix}\hat{\nu}^{\mu}\\ \hat{\mu} \end{pmatrix}_{L}$  \\
	$\hat{\ell}^{\tau}_{L}=\begin{pmatrix}\hat{\nu}^{\tau}\\ \hat{\tau} \end{pmatrix}_{L}$ 
\end{tabular} &
\begin{tabular}{c}
		$0^{+}$	\\
	\\	$0^{+}$	\\
	\\	$-1^{+}$	\\
\end{tabular} &
\begin{tabular}{c}
		\\
	\\		\\
	\\		\\
\end{tabular}   &
\begin{tabular}{c}
	$ \begin{matrix}\hat{\nu}^{e\; c}_{L}	\\ \hat{\nu}^{\mu\; c}_{L} \end{matrix}$  \\
	$ \begin{matrix}\hat{\nu}^{\tau\; c }_{L}	\\  \hat{e}^{e\; c}_{L} \end{matrix}$  \\
	$ \begin{matrix} \hat{e}^{\mu\; c }_{L}	\\  \hat{e}^{\tau\; c}_{L} \end{matrix}$ 
\end{tabular} &
\begin{tabular}{c}
	$ \begin{matrix} \sfrac{-1}{3}^{-}	\\ \sfrac{-1}{3}^{-} \end{matrix}$  \\
	$ \begin{matrix} \sfrac{-1}{3}^{-}	\\ \sfrac{+4}{3}^{-} \end{matrix}$  \\
	$ \begin{matrix} \sfrac{+1}{3}^{-}	\\ \sfrac{+4}{3}^{-} \end{matrix}$ 
\end{tabular} &
\begin{tabular}{c}
	  \\
	 \\
\end{tabular}
\\ \hline\hline 

\multicolumn{6}{c}{Non-SM Quarks: $Y_{\mathcal{T}_{L}}=-Y_{\mathcal{T}_{L}^{c}}=\sfrac{-4}{3}$, $Y_{\mathcal{J}_{L}}=-Y_{\mathcal{J}_{L}^{c}}=\sfrac{+2}{3}$}\\ \hline\hline	
\begin{tabular}{c}	
	$\hat{\mathcal{T}}_{L}$	\\	
	$\mathcal{J}_{L}^{1}$	\\	$\mathcal{J}_{L}^{2}$	
\end{tabular} &
\begin{tabular}{c}
	$\sfrac{+1}{3}^{-} $	\\	
	$ 0^{+} $           	\\	$ 0^{+} $
\end{tabular}   &
\begin{tabular}{c}
		\\		\\	
		\\		
\end{tabular}   &

\begin{tabular}{c}
	$\hat{\mathcal{T}}_{L}^{c}$	\\	
	$\hat{\mathcal{J}}_{L}^{c \ 1}$	\\	$\hat{\mathcal{J}}_{L}^{c \ 2}$	
\end{tabular} &
\begin{tabular}{c}
	$\sfrac{-2}{3}^{-} $	\\	  
	$\sfrac{+1}{3}^{+} $	\\	$\sfrac{+1}{3}^{+} $
\end{tabular} &
\begin{tabular}{c}
		\\		\\	
		\\		
\end{tabular}
\\ \hline\hline

\multicolumn{6}{c}{Non-SM Leptons: $Y_{E_{L}}=-Y_{E_{L}^{c}}=Y_{\mathcal{E}_{L}}=-Y_{\mathcal{E}_{L}^{c}}=-2$}\\ \hline\hline	
\begin{tabular}{c}	
    $\hat{E}_{L}$	\\
    $\hat{\mathcal{E}}_{L}$	\\
\end{tabular} &
\begin{tabular}{c}
	$-1^{+}$	    	\\
	$\sfrac{-2}{3}^{+}$	\\
\end{tabular} &
\begin{tabular}{c}
			\\
			\\
		
\end{tabular}   &
\begin{tabular}{c}	
    $\hat{E}_{L}^{c}$	\\
    $\hat{\mathcal{E}}_{L}^{c}$	\\
\end{tabular} &
\begin{tabular}{c}
	$\sfrac{+2}{3}^{+}$		\\
	$+1^{+}$	            \\
\end{tabular} &
\begin{tabular}{c}
			\\
			\\
	
\end{tabular}
\\	\hline\hline 
\multicolumn{3}{c}{Majorana Fermions: $Y_{\mathcal{N}}=0$} & 
\begin{tabular}{c}	
	$\mathcal{N}_{R}^{1,2,3}$	\\
\end{tabular} &
\begin{tabular}{c}
	$0^{-}$	\\
\end{tabular} &
\begin{tabular}{c}
	\\
\end{tabular}
\\	\hline\hline 
\end{tabular}
\end{table}

The scalar singlets $\sigma$ and $\sigma'$ do not acquire VEV but they contribute to the generation of the lightest fermions masses at one-loop level. The scalar singlets $\chi$, $\chi'$ are responsible of $Z'$ gauge boson and exotic particle masses where the former can be approximated to $M_{Z'}\approx \sfrac{g_{X}\sqrt{v_{\chi}^{2}+v_{\chi}^{\prime2}}}{3}$ so it is reasonable to think that they acquire a VEV at least at the TeV scale, more specifically for a $Z'$ mass above $5$TeV it is required that  $\sqrt{v_{\chi}^{2}+v_{\chi}^{\prime2}}>18$ TeV. These $\chi$, $\chi'$ VEVs break the $U(1)_{X}$ symmetry, leading to the following spontaneous symmetry breaking chain:
\begin{equation}
    \mathrm{SU(3)}_{C}\otimes
    \mathrm{SU(2)}_{L}\otimes 
    \mathrm{U(1)}_{Y} \otimes 
    \mathrm{U(1)}_{X} \overset{\chi}{\longrightarrow}
    \mathrm{SU(3)}_{C}\otimes
    \mathrm{SU(2)}_{L}\otimes 
    \mathrm{U(1)}_{Y} \overset{\Phi}{\longrightarrow}
    \mathrm{SU(3)}_{C}\otimes
    \mathrm{U(1)}_{Q}. \nonumber
\end{equation}

In the context of the supersymmetric theory, the gauge groups induce a D-term potential shown in Eq. (\ref{VD}) and the renormalizable and gauge invariant superpotential given in Eq. (\ref{VW}) divided into scalar ($W_{\phi}$), quark ($W_{Q}$) and lepton ($W_{L}$) parts shown in Eqs. (\ref{4-9}-\ref{4-11})
\begin{small}
\begin{align}
V_{D}&=\frac{g^{2}}{2}\Big[ |\Phi_{1}^{\dagger}\Phi_{2}|^{2}+|\Phi_{1}^{\prime\dagger}\Phi_{2}'|^2+|\Phi_{1}^{\prime\dagger}\Phi_{1}|^2+|\Phi_{1}^{\prime\dagger}\Phi_{2}|^2+|\Phi_{2}^{\prime\dagger}\Phi_{1}|^2+|\Phi_{2}^{\prime\dagger}\Phi_{2}|^2\nonumber\\
 &-|\Phi_{1}|^{2}|\Phi_{2}|^{2}-|\Phi_{1}^{\prime}|^{2}|\Phi_{2}^{\prime}|^{2} \Big]+\frac{g^{2} + g^{\prime 2}}{8}(\Phi_{1}^{\dagger}\Phi_{1}+\Phi_{2}^{\dagger}\Phi_{2}-\Phi_{1}^{\prime\dagger}\Phi_{1}^{\prime}-\Phi_{2}^{\prime\dagger}\Phi_{2}^{\prime})^{2} \nonumber\\
 &+\frac{g_{X}^{2}}{2}\left[\frac{2}{3}(\Phi_{1}^{\dagger}\Phi_{1}-\Phi_{1}^{\prime\dagger}\Phi_{1}^{\prime})+\frac{1}{3}(\Phi_{2}^{\dagger}\Phi_{2}-\Phi_{2}^{\prime\dagger}\Phi_{2}^{\prime})-\frac{1}{3}(\chi^{*}\chi-\chi^{\prime*}\chi^{\prime}) -\frac{1}{3}(\sigma^{*}\sigma-\sigma^{\prime*}\sigma^{\prime})\right]^{2}, \label{VD} \\
W[\Phi]&=W_{\phi} + W_{Q} + W_{L} , \label{VW} \\
    W_{\phi}&=-\mu_{1}\hat{\Phi}'_{1}\hat{\Phi}_{1}-\mu_{2}\hat{\Phi}'_{2}\hat{\Phi}_{2} - \mu_{\chi}\hat{\chi} '\hat{\chi} - \mu_{\sigma}\hat{\sigma} '\hat{\sigma} + \lambda_{1}\hat{\Phi}_{1}^{\prime}\hat{\Phi}_{2}\hat{\sigma}^{\prime} + \lambda_{2}\hat{\Phi}_{2}^{\prime}\hat{\Phi}_{1}\sigma, \label{4-9} \\
    W_{Q}&=\hat{q}_{L}^{1}\hat{\Phi}_{2}h_{2u}^{12}\hat{u}_{L}^{2\; c} + \hat{q}_{L}^{2}\hat{\Phi}_{1}h_{1u}^{22}\hat{u}_{L}^{2\; c} + \hat{q}_{L}^{3}\hat{\Phi}_{1}h_{1u}^{3k}\hat{u}_{L}^{k\; c} - \hat{q}_{L}^{3}\hat{\Phi '}_{2}h_{2d}^{3j}\hat{d}_{L}^{j\; c}+ \hat{q}_{L}^{1}\hat{\Phi}_{2}h_{2T}^{1}\hat{\mathcal T}_{L}^{c}\nonumber\\
    & +\hat{q}_{L}^{2}\hat{\Phi}_{1}h_{1T}^{2}\hat{\mathcal T}_{L}^c-\hat{q}_{L}^{1}\hat{\Phi '}_{1}{h}_{1J}^{1a}\hat{\mathcal J}_{L}^{a\; c} - \hat{q}_{L}^{2}\hat{\Phi '}_{2}{h}_{2J}^{2a}\hat{\mathcal J}_{L}^{a\; c} + \hat{\mathcal T}_{L}\hat{\chi}'h_{\chi'}^{ T}\hat{\mathcal T}_{L}^{c} - \hat{\mathcal J}_{L}^{a}\hat{\chi}h_{\chi J}^{a b}\hat{\mathcal J}_{L}^{b\; c} \nonumber\\
    &+\hat{\mathcal T}_{L}\hat{\chi '}h_{\chi' u}^{T2}\hat{u}_{L}^{2\; c} + \hat{\mathcal{J}}_{L}^{a}\hat{\sigma} h_{\sigma d}^{aj}\hat{d}_{L}^{j\; c } + \hat{\mathcal{T}}_{L}\hat{\sigma}^{\prime}h_{\sigma^{\prime}u}^{Tk}\hat{u}_{L}^{k\; c}, \label{4-10} \\
    W_{L}&= \hat{\ell}_{L}^{p}\hat{\Phi}_{2}h_{2\nu}^{pq}\hat{\nu}_{L}^{q\; c} -\hat{\ell}_{L}^{p}\hat{\Phi}'_{2}{h}_{2e}^{p\mu}\hat{e}_{L}^{\mu\; c}
    - \hat{\ell}_{L}^{\tau}\hat{\Phi}'_{2}{h}_{2e}^{\tau r}\hat{e}_{L}^{r\; c} 
    - \hat{\ell}_{L}^{p}\hat{\Phi}'_{1}{h}_{1E}^{p}\hat{E}_{L}^{c}+ \hat{E}_{L}\hat{\chi}'{g}_{\chi' E}\hat{E}_{L}^{c}  \nonumber \\
    &- \hat{E}_{L}\mu_{E}\hat{\mathcal{E}}_{L}^{c} + \hat{\mathcal{E}}_{L}\hat{\chi}g_{\chi\mathcal{E}}\hat{\mathcal{E}}_{L}^{c}  - \hat{\mathcal{E}}_L\mu_{\mathcal{E}}\hat{E}_{L}^{c} +\hat{\nu}_{L}^{q\; c}\hat{\chi}' {h}_{\chi N}^{\prime\; qn}\hat{N}_{L}^{n\; c}
        + \frac{1}{2}\hat{N}_{L}^{m\; c} M_{mn}\hat{N}_{L}^{n\; c} \nonumber\\
        &+ \hat{E}_{L}\hat{\sigma} h_{\sigma e}^{E r}\hat{e}_{L}^{c\; r} + \hat{\mathcal{E}}_{L}\hat{\sigma}^{\prime}h_{\sigma' e}^{\mathcal{E} \mu}\hat{e}_{L}^{\mu\; c},\label{4-11}
\end{align}
\end{small}

\noindent where $j=1,2,3$ labels the MSSM down-like singlet quark superfields, $k=1,3$ labels the first and third generation of singlet up-like quark superfields, $a=1,2$ is the index of the new $\mathcal{J}_{L}^{a}$ and $\mathcal{J}_{L}^{c a}$ singlet down-like quark superfields, $p=e,\mu$ labels the first and second lepton doublet superfields, $q=e,\mu,\tau$ labels the neutral lepton superfields containing right-handed neutrinos, $r=e,\tau$ is the index of the MSSM right-handed charged lepton superfields and $m,n=1,2,3$ label the neutral lepton superfields containing Majorana neutrinos. All of them according to the notation given in table \ref{modelfermions}.

Finally, the relevant soft breaking potential is included. Since we are not interested in sparticle masses let's consider only soft breaking terms for the scalar particles and gauginos as shown in Eq. (\ref{Vs}) bellow
\begin{align}\label{Vs}
    V_{soft}&=m_{1}^{2}\Phi_{1}^{\dagger}\Phi_{1} + {m}_{1}^{\prime 2}{\Phi}_{ 1}^{\prime \dagger}\Phi'_{1} + m_{2}^{2}\Phi_{2}^{\dagger}\Phi_{2} + {m}_{2}^{\prime 2}\Phi _{2}^{\prime\dagger}\Phi'_{2}+m_{\chi}^{2}\chi^{\dagger}\chi + {m}_{\chi}^{\prime 2}{\chi}^{\prime\dagger}\chi'+m_{\sigma}^{2}\sigma^{\dagger}\sigma \nonumber\\
    & + {m}_{\sigma}^{\prime 2}{\sigma}^{\prime\dagger}\sigma'  -\bigg[\mu_{11}^{2}\epsilon_{ij}({\Phi}_{1}^{\prime i}\Phi_{1}^{j}) -\mu_{22}^{2}\epsilon_{ij}({\Phi}_{2}^{\prime i}\Phi_{2}^{j}) -\mu_{\chi\chi}^{2}(\chi\chi') +\mu_{\sigma\sigma}^{2}(\sigma\sigma') + \tilde{\lambda}_{1}\Phi_{1}^{\prime \dagger}\Phi_{2}\sigma^{\prime}\nonumber\\
    & + \tilde{\lambda}_{2}\Phi_{2}^{\prime \dagger}\Phi_{1}\sigma - \frac{2\sqrt{2}}{9}(k_{1}\Phi_{1}^{\dagger}\Phi_{2}\chi' -k_{2}\Phi_{1}^{\dagger}\Phi_{2}\chi^*+k_{3}\Phi_{1}'{}^{\dagger}\Phi_{2}'\chi -k_{4}\Phi_{1}'{}^{\dagger}\Phi_{2}'\chi'{}^*)+h.c.\bigg] \nonumber \\
    &+ M_{\tilde{B}}\tilde{B}\tilde{B}^{\dagger} + M_{\tilde{B}'}\tilde{B}'\tilde{B}^{\prime \dagger} + M_{\tilde{W}^{\pm}}\tilde{W}^{\pm}\tilde{W}^{\pm \dagger} + M_{\tilde{W}}\tilde{W}_{3}\tilde{W}_{3}^{\dagger},
\end{align}
where the last terms, proportional to the coupling constants named $k_{1},k_{2},k_{3}$ and $k_{4}$, also breaks softly the parity symmetry to avoid scalar particles below the $125.3$ GeV threshold. Although  F-term potential codifies mainly all sparticles interactions and off-diagonal sparticle mass terms, we take only the associated with Higgs particles:
\begin{eqnarray}\label{VF}
V_{F}&=&\mu_{1}^2 ( \Phi_{1}^\dagger\Phi_{1}+\Phi_{1}^{\prime\dagger}\Phi_{1}^{\prime})+\mu_{2}^2(\Phi_{2}^\dagger\Phi_{2}+\Phi_{2}^{\prime\dagger}\Phi_{2}^{\prime})
+\mu_{\chi}^2(\chi^*\chi+\chi^{\prime*}\chi') + +\mu_{\sigma}^2(\sigma^*\sigma+\sigma^{\prime*}\sigma^{\prime})\nonumber\\
&+& (\lambda_{1}^{2}|\epsilon_{ij}\Phi_{1}^{\prime i}\Phi_{2}^{j}|^{2} + \lambda_{2}^{2}|\epsilon_{ij}\Phi_{2}^{\prime i}\Phi_{1}^{j}|^{2}  + \lambda_{1}^{2}( \Phi_{2}^{\dagger} \Phi_{2} + \Phi_{1}^{\prime \dagger}\Phi_{1}^{\prime}\sigma^{\prime *}\sigma^{\prime} + \lambda_{2}^{2}( \Phi_{1}^{\dagger}\Phi_{1} + \Phi_{2}^{\prime \dagger}\Phi_{2}^{\prime})\sigma^{*}\sigma \nonumber\\
& -&\lambda_{1}\mu_{1}\Phi_{1}^{\dagger}\Phi_{2}\sigma^{\prime} - \lambda_{1}\mu_{2}\Phi_{2}^{\prime \dagger }\Phi_{1}^{\prime}\sigma^{\prime}  -\lambda_{2}\mu_{1}\Phi_{1}^{\prime \dagger}\Phi_{2}^{\prime}\sigma -\lambda_{2}\mu_{2}\Phi_{2}^{\dagger}\Phi_{1}\sigma - \lambda_{1}\mu_{\sigma}\epsilon_{ij}\Phi_{1}^{\prime i}\Phi_{2}^{j} \nonumber\\
& -&\lambda_{2}\mu_{\sigma}\epsilon_{ij}\Phi_{2}^{\prime i }\Phi_{1}^{j} + h.c. ).
\end{eqnarray}
It is worth to mention that only the terms involving interactions among Higgs bosons have been considered in the $V_{D}$ potential since we are not interested in the sparticles mass generation or interactions.

\section{Lepton Masses}\label{leptonsector}
\subsubsection{Charged leptons masses and 1-loop corrections}
\begin{figure}[H]
    \centering
    \includegraphics[scale=0.2]{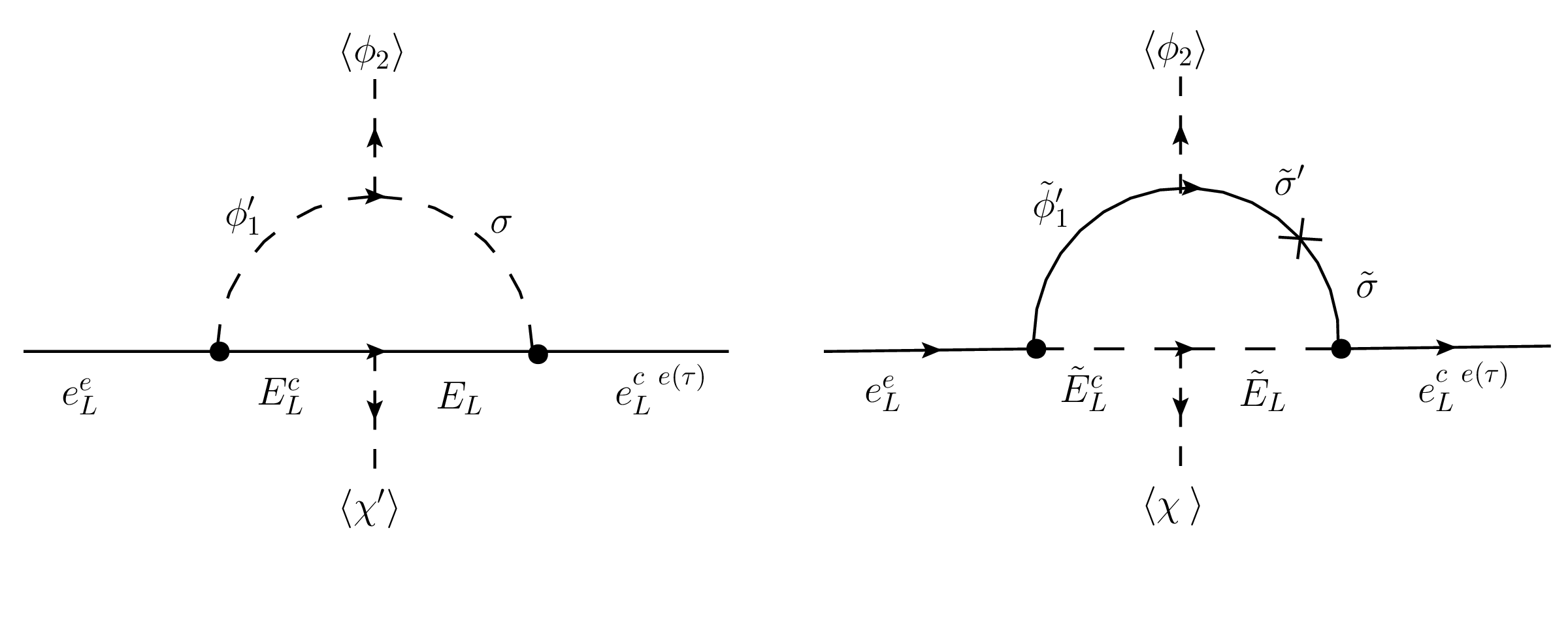}
    \caption{One-loop corrections to the leptons due to exotic fermions, sfermions and Higgsinos.}
    \label{fig:1-loopforleptons}
\end{figure}

\noindent
From the lepton super potential in Eq. (\ref{4-11}) we can write the following mass matrix in the basis $(e,\mu.\tau,E,\mathcal{E})$:
\begin{align}
    \mathcal{M}_{E}&=\frac{1}{\sqrt{2}}\left(\begin{array}{ c c c |c c}
    0                           & h_{2e}^{e\mu}v'_{2}     & 0 &  h_{1E}^{e}v'_{1}    & 0 \\
    0                           & h_{2e}^{\mu\mu}v'_{2}   & 0 &  h_{1E}^{\mu}v'_{1}  & 0 \\
    h_{2e}^{\tau e}v'_{2}  & 0                            & h_{2e}^{\tau\tau}v'_{2} & 0 & 0 \\ \hline
    0 & 0 & 0 & {g}_{\chi' E}v'_{\chi} & -\mu_{E} \\
    0 & 0 & 0 & -\mu_{\mathcal{E}} & g_{\chi\mathcal{E}}v_{\chi}  \\
    \end{array} \right).
\end{align}
\noindent
Such matrix has two exotic lepton singlets coupled by $\mu_{E}$ and $\mu_{\mathcal{E}}$ but the $\mathcal{E}$ fermion do not mix with any of the SM particles so we consider the decoupled case $\mu_{E}=\mu_{\mathcal{E}}=0$. Furthermore, the squared mass matrix $\mathcal{M}_{E}\mathcal{M}_{E}^{\dagger}$ has rank $4$, then the electron turns out to be massless and radiative corrections must be considered. Figure \ref{fig:1-loopforleptons} shows the diagrams that contribute to the electron mass at one-loop level, inducing new terms in the $(1,1)$ and $(1,3)$ entries in the mass matrix, which now reads:

\begin{align}
    \mathcal{M}_{E}^{1-Loop}&=\frac{1}{\sqrt{2}}\left(\begin{array}{ c c c |c c}
    v_{2}\Sigma_{ee}                           & h_{2e}^{e\mu}v'_{2}     & v_{2}\Sigma_{e\tau} &  h_{1E}^{e}v'_{1}    & 0 \\
    0                           & h_{2e}^{\mu\mu}v'_{2}   & 0 &  h_{1E}^{\mu}v'_{1}  & 0 \\
    h_{2e}^{\tau e}v'_{2}  & 0                            & h_{2e}^{\tau\tau}v'_{2} & 0 & 0 \\ \hline
    0 & 0 & 0 & {g}_{\chi' E}v'_{\chi} & -\mu_{E} \\
    0 & 0 & 0 & -\mu_{\mathcal{E}} & g_{\chi\mathcal{E}}v_{\chi}  \\
    \end{array} \right).
\end{align}

\noindent
The radiative corrections can be done thanks to the interactions terms given by:
\begin{align}
    W_{\phi}&\Rightarrow\lambda_{1}\hat{\Phi}_{1}^{\prime}\hat{\Phi}_{2}\hat{\sigma}^{\prime} -\mu_{\sigma}\hat{\sigma}^{\prime}\hat{\sigma},&
    W_{L}&\Rightarrow\hat{E}_{L}\hat{\sigma} h_{\sigma e}^{E r}\hat{e}_{L}^{r \;c} + \hat{\ell}_{L}^{e}\hat{\Phi '_{1}}{h}_{1E}^{e}\hat{E}_{L}^{c} + \hat{E}_{L}\hat{\chi '}{g}_{\chi' E}\hat{E}_{L}^{c}, 
\end{align}
where $r=e,\tau$, the couplings $\lambda_{1}$,  $h_{\sigma e}^{E r}$, $g_{\chi' E}$ and $h_{1E}^{e}$ are dimensionless Yukawa couplings and $\mu_{\chi}$ and $\mu_{\sigma}$ are mass unit parameters from the scalar potential. The first diagram in figure \ref{fig:1-loopforleptons} illustrates the non-SUSY contribution which is given by:
\begin{align}
    v_{2}\Sigma _{ee(e\tau)}^{NS}=v_{2} h_{\sigma e}^{E e(\tau)} \tilde{\Sigma}^{NS}=\frac{-1}{16\pi ^2}\frac{v_{2}}{\sqrt{2}}\frac{\lambda_{1}\mu_{\sigma}h_{\sigma e}^{E e(\tau)}h_{1E}^e}{M_{E}}C_0\left(\frac{m_{h1}^{\prime}}{M_{E}},\frac{m_{\sigma}^{\prime}}{M_E}\right),
\end{align}
where $M_{E}$ is the exotic charged fermion mass, $m_{h1}^{\prime}$ is the corresponding mass of the $h'_{1}$ field in flavor basis just like  $m_{\sigma}^{\prime}$ is for the $\sigma$ field and $C_{0}$ is the Veltmann-Passarino function evaluated for $p^{2}=0$ given by eq. (\ref{C0})\cite{tablaC0} and the double fermion propagator in the SUSY contribution makes the diagram to have a term proportional to the Passarino Veltman-Function $C_{00}$\cite{Scalarintegralshighenergy}  that can be decomposed in terms of the scalar integrals $C_{0}$ and $B_{0}$ \cite{PVscalarfunctions}.On the other hand, the SUSY contribution is given by:
\begin{small}
\begin{align}
    v_{2}\Sigma_{ee(e\tau)}^{S}&=v_{2} h_{\sigma e}^{E e(\tau)} \tilde{\Sigma}^{S} = -\frac{1}{32\pi^{2}}\frac{v_{2}}{\sqrt{2}}\sum_{n=1}^{10}\sum_{k=1}^{2}Z_{L}^{9n}Z_{L}^{4n} Z_{\tilde{h}}^{10 k}Z_{\tilde{h}}^{11 k} \lambda_{1}\mu_{\sigma}h_{\sigma e}^{E e(\tau)}h_{1E}^e\times \\
    &\times  \left[\frac{(\tilde{m}_{\sigma  k}+\tilde{m}_{h_{1}}^{\prime})^{2}}{\tilde{M}_{L_{n}}^{2}}C_{0}\left(\frac{\tilde{m}_{h1}^{\prime}}{\tilde{M}_{L_{n}}},\frac{\tilde{m}_{\sigma  k}}{\tilde{M}_{L_{n}}} \right) + \tilde{m}_{h1}^{\prime 2}B_{0}(0,\tilde{m}_{\sigma}^{\prime},\tilde{M}_{L_{n}}) + \tilde{m}_{\sigma k}^{2}B_{0}(0,\tilde{m}_{h1}^{\prime},\tilde{M}_{L_{n}}) \right], \nonumber \\
    \nonumber
    \end{align}
\noindent
where
    \begin{align}
    C_{0}(m_{1},m_{2}) &= \frac{1}{(1-m_{1}^{2})(1-m_{2}^{2})(m_{1}^{2}-m_{2}^{2})}\left[m_{1}^{2}m_{2}^{2}\ln\left(\frac{m_{1}^{2}}{m_{2}^{2}}\right) + m_{2}^{2}\ln(m_{2}^{2})- m_{1}^{2}\ln(m_{1}^{2})\right], \label{C0}
\end{align}
\end{small}
$\tilde{M}_{L_{n}}$ are the charged sleptons mass eigenvalues, $Z_{\tilde{h}}$ is the rotation matrix that connects $\tilde{\sigma}$ ($\tilde{\sigma'}$)  with its mass eigenstates with eigenvalues $\tilde{m}_{h k}$ which are running inside the loop. In this case, higgsinos are organized in the basis $(\tilde{B},\tilde{W}_{3},\tilde{B}',\tilde{h}_{1},\tilde{h}'_{1},\tilde{h}_{2},\tilde{h}'_{2},\tilde{\chi},\tilde{\chi'},\tilde{\sigma}, \tilde{\sigma'} )$. Likewise, sleptons are assumed in the basis $(\tilde{e}_{L},\tilde{\mu}_{L},\tilde{\tau}_{L},\tilde{E}_{L},\tilde{\mathcal{E}}_{L},\tilde{e}_{L}^{c},\tilde{\mu}_{L}^{c},\tilde{\tau}_{L}^{c},\tilde{E}_{L}^{c},\tilde{\mathcal{E}}_{L}^{c})$ so the sleptons inside the loop are identified with the 4th and 9th states and $Z_{L}$ is the rotation matrix that connects the exotic sleptons their respective eigenstates $\tilde{L}_{n}$ inside the loop. The final expressions for mass eigenvalues are given by:
\begin{align}
m_{e}^{2}&=\frac{1}{2}v_{2}^{2}\tilde{\Sigma}^{2}\frac{(h_{\sigma e}^{E e} 
h_{2 e}^{\tau \tau } - h_{\sigma e}^{E\tau}h_{2 e}^{\tau e})^{2}}{(h_{2e}^{\tau e})^2+(h_{2e}^{\tau \tau})^2}, & m_{\mu}^{2}&=\frac{1}{2}v_2'{}^{2}\left[(h_{2e}^{e\mu})^2+(h_{2e}^{\mu\mu})^2\right], \nonumber\\
m_{\tau}^{2}&=\frac{1}{2}v_2'{}^2
\left[(h_{2e}^{\tau e})^2+(h_{2e}^{\tau \tau})^2\right], & m_{E}^{2}&=\frac{1}{2}g_{\chi' E}^2\; v_\chi'{}^2, \nonumber\\
m_{\mathcal{E}}^{2}&=\frac{1}{2}g_{\chi \mathcal{E}}^2\; v_\chi^2, \label{CLmasses}
 \end{align}
where
\begin{align}
    \tilde{\Sigma}&=\tilde{\Sigma}^{NS}+\tilde{\Sigma}^{S}.
\end{align}

\noindent
To provide simplified expressions, we define a $\theta_{e\tau}$ angle according to:

\begin{align}
    \tan \theta_{e\tau} &= t_{e\tau} \equiv \frac{h_{2e}^{\tau e}}{h_{2e}^{\tau\tau}}, & \sqrt{(h_{2e}^{\tau e})^{2}+ (h_{2e}^{\tau\tau})^{2}} &= \frac{\sqrt{2}m_{\tau}}{v'_{2}},
\end{align}
\noindent
which determines $h_{2e}^{\tau e}$ and $h_{2e}^{\tau\tau}$ for a given $\theta_{e\tau}$ angle. Likewise, $h_{2e}^{e\mu}$ and $h_{2e}^{\mu\mu}$ are obtained given a $\theta_{e\mu}$ angle according to: 

    \begin{align}
    \tan \theta_{e\mu} &= t_{e\mu} \equiv \frac{h_{2e}^{e\mu}}{h_{2e}^{\mu\mu}}, & \sqrt{(h_{2e}^{e\mu})^{2}+ (h_{2e}^{\mu\mu})^{2}} &= \frac{\sqrt{2}m_{\mu}}{v'_{2}},
\end{align}

Besides, we define a set of rotated parameters given by:

\begin{align}
    \begin{pmatrix}
    \Omega_{1} \\
    \Omega_{2}
    \end{pmatrix}
    =\begin{pmatrix}
    c_{e\tau} & -s_{e\tau} \\
    s_{e\tau} & c_{e\tau}
    \end{pmatrix}
    \begin{pmatrix}
    h_{\sigma e}^{E e}\\
    h_{\sigma e}^{E \tau}
    \end{pmatrix}.
\end{align}

Such definition allow us to write the electron mass as $ m_{e}^{2}=\frac{1}{2}v_{2}^{2}\tilde{\Sigma}^{2}\Omega_{1}^{2} $
and the rotation matrices as $V^{\ell}_{L}=V_{2L}^{\ell}V_{1L}^{\ell}$ being each matrix defined by:
\small
\begin{align}\label{v1l}
    V_{1L}^{\ell}&=\left(\begin{array}{ccc|cc}
      1 & 0 & 0 & -r_{e\chi 1} & -r_{e\chi 2} \\
      0 & 1 & 0 & -r_{e\chi 3} & -r_{e\chi 4} \\
      0 & 0 & 1 & 0 &0 \\ \hline
     r_{e\chi 1} &  r_{e\chi 3} & 0 & 1 &0 \\
     r_{e\chi 2} &r_{e\chi 4} & 0 & 0 &1 
    \end{array} \right), 
&
V_{2L}^{\ell}&=\left(\begin{array}{ccc|cc}
 \cos \theta_{e\mu} & \sin \theta_{e\mu} & -\frac{v_{2}}{v'_{2}}\Omega_{2}\tilde{\Sigma}\cos \theta_{e\mu}  & 0 & 0 \\
 -\sin \theta_{e\mu} & \cos \theta_{e\mu} & \frac{v_{2}}{v'_{2}}\Omega_{2}\tilde{\Sigma}\sin \theta_{e\mu}   & 0 & 0 \\
 \frac{v_{2}}{v'_{2}}\Omega_{2}\tilde{\Sigma} & 0 & 1  & 0 & 0 \\
 \hline
 0 & 0 & 0 & 1 & 0 \\
 0 & 0 & 0 & 0 & 1 \\
\end{array} \right),
\end{align}

\normalsize
where
\begin{align}
r_{e\chi 1}&\approx \frac{h_{1e}^E v'_1}{\sqrt{2}m_{E}}, & r_{e\chi 2}&\approx \frac{h_{1e}^E \mu_{E} v'_1}{2m_{E}m_{\mathcal{E}}}, \nonumber \\
r_{e\chi 3}& \approx \frac{h_{1\mu}^E v'_1}{\sqrt{2}m_{E}}, & r_{e\chi 4}& \approx \frac{h_{1\mu}^E \mu_{E} v'_1}{2m_{E}m_{\mathcal{E}}} ,
\end{align}
and $V_{1L}^{\ell}$ is a matrix that diagonalizes the exotic leptons via seesaw mechanism and $V_{2L}^{\ell}$ diagonalizes the SM leptons being the mixing angle $\theta_{e\mu}$.

Then, the total rotation matrix for charged leptons can be written as:
\footnotesize
\begin{align}\label{VL}
    V^{\ell}_{L}&=V_{2L}^{\ell}V_{1L}^{\ell} \nonumber \\
    &=\left(\begin{array}{ccc|cc}
       \cos\theta_{e\mu} & \sin\theta_{e\mu} & - \frac{v_{2}}{v'_{2}}\Omega_{2}\tilde{\Sigma}\cos\theta_{e\mu} & -\sin\theta_{e\mu} r_{e\chi 3}-r_{e\chi 1} \cos\theta_{e\mu} & -\sin\theta_{e\mu} r_{e\chi 4}-r_{e\chi 2} \cos\theta_{e\mu} \\
 -\sin\theta_{e\mu} & \cos\theta_{e\mu} & \frac{v_{2}}{v'_{2}}\Omega_{2}\tilde{\Sigma}\sin\theta_{e\mu} & \sin\theta_{e\mu} r_{e\chi 1}-\cos\theta_{e\mu} r_{e\chi 3} & \sin\theta_{e\mu} r_{e\chi 2}-\cos\theta_{e\mu} r_{e\chi 4} \\
 \frac{v_{2}}{v'_{2}}\Omega_{2}\tilde{\Sigma} & 0 & 1 & -\frac{v_{2}}{v'_{2}}\Omega_{2}\tilde{\Sigma}r_{e\chi 1}  & -\frac{v_{2}}{v'_{2}}\Omega_{2}\tilde{\Sigma} r_{e\chi 2}  \\ \hline
 r_{e\chi 1} & r_{e\chi 3} & 0 & 1 & 0 \\
 r_{e\chi 2} & r_{e\chi 4} & 0 & 0 & 1 \\
    \end{array} \right).
\end{align}
\normalsize

 In addition, the $v_{2}^{\prime}$ VEV is related to muon and tau lepton masses so an estimate of some couplings can be done by considering the physical mass ratio of $\mu$ and $\tau$ masses which  is approximately $\frac{m_{\mu}}{m_{\tau}}\approx 0.14$, then:
\begin{align}
  \frac{m_{\mu}}{m_{\tau}} =  0.14 \approx \frac{\sqrt{(h_{2e}^{e\mu})^2+(h_{2e}^{\mu\mu})^2}}{\sqrt{(h_{2e}^{\tau e})^2+(h_{2e}^{\tau \tau})^2}} =\frac{h_{2e}^{\mu\mu}}{h_{2e}^{\tau\tau}}\frac{c_{e\tau}}{c_{e\mu}}.\label{muovertau}
\end{align}
In particular, we can consider the case $h_{2e}^{\mu \mu} \approx h_{2e}^{\tau \tau}$ Eq. (\ref{muovertau}) provides a relationship between $\theta_{e\mu}$ and $\theta_{e\tau}$ given by:
\begin{align}\label{emuetau}
    c_{e\tau}=\frac{m_{\mu}}{m_{\tau}}c_{e\mu}
\end{align}

\subsubsection{Neutrino masses at tree-level}

The neutral lepton sector of the model considers additional right-handed and Majorana neutrinos, which according to the interactions considered in Eq. (\ref{4-11}) leads to a mass matrix in the basis $(\nu_{L}^q,\nu_{L}^q{}^{C},N_{L}^{i}{}^C)$ as follows:
\begin{align}
\mathcal{M}_{\nu} &=\begin{pmatrix}
    0 & m_{D}^{T} & 0 \\
    m_{D} & 0 & M_{D}^{T} \\
    0 & M_{D} & M_{M}
    \end{pmatrix},     
\end{align}
where the block matrices are given by:
\begin{align}
 &m_{D}=\frac{v_{2}}{\sqrt{2}}\begin{pmatrix}
    h_{2\nu}^{ee} & h_{2\nu}^{e\mu} & h_{2\nu}^{e\tau} \\
    h_{2\nu}^{e\mu} & h_{2\nu}^{\mu\mu} & h_{2\nu}^{\mu\tau} \\
    0 & 0 & 0
    \end{pmatrix},\ \ \
     (M_{D})^{ij}=\frac{v'_{\chi}}{\sqrt{2}}{h}_{\chi N}^{\prime\; ij}, \ \ \ \ \  (M_{M})_{ij}=\frac{1}{2}M_{ij}.
\end{align}

Such mass matrix can be diagonalized by the Inverse See Saw mechanism (ISS) if we assume the hierarchy $M_{M}\ll m_{D} \ll M_{D} $ \cite{inverseseesaw}. Therefore, block diagonalization is done by the rotation matrix $\mathbb{V}_{SS}$:
\begin{align}
\mathbb{V}_{SS}\mathcal{M}_{\nu}\mathbb{V}_{SS}^{\dagger}
&\approx 
\begin{pmatrix}
m_{light}&0\\
0&m_{heavy}
\end{pmatrix},
\end{align}
resulting in a $3 \times 3$ matrix $m_{light}$ for active neutrinos and a $6 \times 6$ mass matrix $m_{heavy}$ for the heavy Majorana neutrinos, such matrices are given by:

\begin{align}
\mathbb{V}_{SS}&=
\begin{pmatrix}
I&-\Theta_{\nu}\\
\Theta_{\nu}^{T}&I
\end{pmatrix},  \;\;\;\;\;\;\;\;\;\;
\Theta_{\nu}=\begin{pmatrix}
0&M_{D}^{T}\\
M_{D}&M_{M}
\end{pmatrix}^{-1}\begin{pmatrix}
m_{D}\\0
\end{pmatrix},
\end{align}

\begin{align}
    m_{light}&\approx m_{D}^{T}(M_{D}^{T})^{-1}M_{M}(M_{D})^{-1}m_{D} ,
\end{align}
and 

\begin{align}
m_{heavy}\approx\begin{pmatrix}0&M_{D}^{T}\\
M_{D}&M_{M}
\end{pmatrix}.  
\end{align}

\noindent
For simplicity and thinking in exotic neutrinos whose masses are big, we can take the particular case where  $M_{D}$ is diagonal and $M_{M}$ is proportional to the identity to explore one of the possible scenarios of the model, i.e.
\begin{align}
M_{D} &= \frac{v_{\chi}}{\sqrt{2}} \left( \begin{matrix}
h_{N\chi 1}	&	0	&	0	\\	0	&	h_{N\chi 2}	&	0	\\	0	&	0	&	h_{\chi N 3}
\end{matrix} \right), &
M_{M} &= \mu_{N} \mathbb{I}_{3\times 3}.
\end{align}
In this way, the light neutrino mass matrix takes the form
\begin{equation}
m_{\mathrm{light}} = \frac{\mu_{N} v_{2}^{2}}{{h_{N\chi 1}}^{2}v_{\chi}^{2}}
\left( 
\begin{matrix}
	\left( h_{2e}^{\nu e}\right)^{2} + \left( h_{2\mu}^{\nu e} \right)^{2} \rho^{2} &
	{h_{2e}^{\nu e}}\,{h_{2e}^{\nu \mu}} + {h_{2\mu}^{\nu e}}\,{h_{2\mu}^{\nu \mu}}\rho^2 	&
	{h_{2e}^{\nu e}}\,{h_{2e}^{\nu \tau}}+ {h_{2\mu}^{\nu e}}\,{h_{2\mu}^{\nu \tau}}\rho^2 	\\
	{h_{2e}^{\nu e}}\,{h_{2e}^{\nu \mu}} + {h_{2\mu}^{\nu e}}\,{h_{2\mu}^{\nu \mu}}\rho^2	&	
	\left( h_{2e}^{\nu \mu} \right)^{2} + \left( h_{2\mu}^{\nu\mu} \right)^{2} \rho^{2}	&	
	{h_{2e}^{\nu \mu}}\,{h_{2e}^{\nu \tau}}+ {h_{2\mu}^{\nu \mu}}\,{h_{2\mu}^{\nu \tau}}\rho^2	\\
	{h_{2e}^{\nu e}}  \,{h_{2e}^{\nu \tau}}+ {h_{2\mu}^{\nu e}}  \,{h_{2\mu}^{\nu \tau}}\rho^2	&	
	{h_{2e}^{\nu \mu}}\,{h_{2e}^{\nu \tau}}+ {h_{2\mu}^{\nu \mu}}\,{h_{2\mu}^{\nu \tau}}\rho^2	&	
	\left( h_{2e}^{\nu \tau} \right)^{2} + \left( h_{2\mu}^{\nu \tau} \right)^{2} \rho^{2}
\end{matrix} \right),
\end{equation}
where we consider $\rho={h_{N\chi 1}}/{h_{N\chi 2}}\approx
 1$ to have nearly degenerate neutrino masses. Next, we see that the light neutrino mass matrix in Eq. (\ref{mlight}) has an overall factor $\frac{\mu_{N} v_{2}^{2}}{{h_{N\chi 1}}^{2}v_{\chi}^{2}}$ which determine the energy scale of neutrino masses, and all $h_{2\mu}^{\nu k}$ terms have a $\rho$ factor. Therefore, to simplify the numerical routine a rescaling of the yukawa couplings was done according to: 

\begin{align}
    \sqrt{\frac{\mu_{N} v_{2}^{2}}{{h_{N\chi 1}}^{2}v_{\chi}^{2}}} h_{2e(\mu)}^{\nu i} &\equiv  \tilde{h}_{2e(\mu)}^{\nu i}, \label{hrescaled}
 \end{align}

Besides, the matrix $m_{\mathrm{light}}$ has null determinant for every possible choice of $M_{D}$ and $M_{M}$ since $m_{D}$ has a row full of zeros, obtaining one massless neutrino. From the characteristic polynomial for $m_{light}m_{light}^{\dagger}$  we can get an expression for the squared mass eigenvalues:
\begin{align}
    m_{\nu_{1}}^{2}&=0, &
    m_{\nu_{2}}^{2}&= \frac{A-\sqrt{A^2 -4B}}{2}, &
    m_{\nu_{3}}^{2}&= \frac{A+\sqrt{A^2 -4B}}{2}, \label{NOmasses}\\
        m_{\nu_{1}}^{2}&= \frac{A+\sqrt{A^2 -4B}}{2}, &
    m_{\nu_{2}}^{2}&= \frac{A-\sqrt{A^2 -4B}}{2}, &
    m_{\nu_{3}}^{2}&=0 \label{IOmasses},
\end{align}
where 
\begin{align}
      A=&2 |\tilde{h}_{2 e}^{\text{$\nu $e}} \tilde{h}_{2 e}^{\nu \mu }+ \tilde{h}_{2 \mu }^{\text{$\nu $e}} \tilde{h}_{2 \mu }^{\nu \mu }|^{2 }+2 |\tilde{h}_{2 e}^{\text{$\nu $e}} \tilde{h}_{2 e}^{\nu \tau }+ \tilde{h}_{2 \mu }^{\text{$\nu $e}} \tilde{h}_{2 \mu }^{\nu \tau }|^{2}\nonumber \\
      &+ |\left(\tilde{h}_{2 e}^{\text{$\nu $e}}\right)^2 + \left(\tilde{h}_{2 \mu }^{\text{$\nu $e}}\right)^2|^{2}+2 |\tilde{h}_{2 e}^{\nu \mu } \tilde{h}_{2 e}^{\nu \tau }+ \tilde{h}_{2 \mu }^{\nu \mu } \tilde{h}_{2 \mu }^{\nu \tau }|^{2} \nonumber \\
      &+|\left(\tilde{h}_{2 e}^{\nu \mu }\right)^2+ \left(\tilde{h}_{2 \mu }^{\nu \mu }\right)^2|^{2} +|\left(\tilde{h}_{2 e}^{\nu \tau }\right)^2+ \left(\tilde{h}_{2 \mu }^{\nu \tau }\right)^2|^{2}, \nonumber \\
     B=& (\tilde{h}_{2 e}^{\nu \tau } \left( \tilde{h}_{2 e}^{\nu \tau *}\left(\tilde{h}_{2 \mu }^{\text{$\nu $e}} \tilde{h}_{2 \mu }^{\text{$\nu $e}*}+\tilde{h}_{2 \mu }^{\nu \mu }  \tilde{h}_{2 \mu }^{\nu \mu *}\right)-  \tilde{h}_{2 \mu }^{\nu \tau *} \left(\tilde{h}_{2 e}^{\text{$\nu $e}*} \tilde{h}_{2 \mu }^{\text{$\nu $e}}+\tilde{h}_{2 e}^{\nu \mu *} \tilde{h}_{2 \mu }^{\nu \mu }\right)\right) \nonumber \\
     & +\tilde{h}_{2 e}^{\text{$\nu $e}} \left(\tilde{h}_{2 \mu }^{\nu \mu } \left(\tilde{h}_{2 e}^{\text{$\nu $e}*} \tilde{h}_{2 \mu }^{\nu \mu *}-\tilde{h}_{2 e}^{\nu \mu *} \tilde{h}_{2 \mu }^{\text{$\nu $e}*}\right)+\tilde{h}_{2 \mu }^{\nu \tau } \left(\tilde{h}_{2 e}^{\text{$\nu $e}*} \tilde{h}_{2 \mu }^{\nu \tau *}-\tilde{h}_{2 e}^{\nu \tau *} \tilde{h}_{2 \mu }^{\text{$\nu $e}*}\right)\right)\nonumber \\
     &+\tilde{h}_{2 e}^{\nu \mu } \left(\tilde{h}_{2 \mu }^{\text{$\nu $e}} \left(\tilde{h}_{2 e}^{\nu \mu *} \tilde{h}_{2 \mu }^{\text{$\nu $e}*}-\tilde{h}_{2 e}^{\text{$\nu $e}*} \tilde{h}_{2 \mu }^{\nu \mu *}\right)+\tilde{h}_{2 \mu }^{\nu \tau } \left(\tilde{h}_{2 e}^{\nu \mu *} \tilde{h}_{2 \mu }^{\nu \tau *}-\tilde{h}_{2 e}^{\nu \tau *} \tilde{h}_{2 \mu }^{\nu \mu *}\right)\right))^{2}.    
\end{align}

The mass spectrum has been written in two different forms because Eq. (\ref{NOmasses}) corresponds to normal ordering while Eq. (\ref{IOmasses}) to inverse ordering. Despite we do not know the mass values, they must be in agreement to squared mass differences in table \ref{tablenu}. For normal ordering we can consider that $m_{3}\gg m_{2},m_{1}$. Therefore we assume $A^{2} \gg 4B$ which let us approximate masses as indicated in Eq. (\ref{normalordering}). In the case of inverse ordering $m_{\nu 1} \approx m_{\nu 2} \approx \frac{A}{2}$ both with a correction $\Delta = \sqrt{A^{2}-4B}$ resulting in the masses of Eq. (\ref{inverseordering}). Finally, the squared mass differences are stated in table \ref{massdifderences}.
\begin{align}
    NO &&m_{\nu_{1}}^{2}&=0, &
    m_{\nu_{2}}^{2}&\approx \frac{B}{A}, &
    m_{\nu_{3}}^{2}&\approx \left( A - \frac{B}{A}\right), \label{normalordering}\\
     IO && m_{\nu_{1}}^{2}&= \frac{A-\Delta}{2},  &
    m_{\nu_{2}}^{2}&= \frac{A+\Delta}{2}, &
    m_{\nu_{3}}^{2}&=0.  \label{inverseordering}
\end{align}

\begin{table}[H]
\centering
\begin{tabular}{|c|c|c|} \hline
   &  Normal Ordering & Inverse Ordering \\ \hline
        $\frac{\Delta m_{21}^{2}}{10^{-5}eV^{2}}$ & $ \frac{B}{A} = 7.39_{-0.20}^{+0.21}$ & $\Delta \approx 7.39_{-0.20}^{+0.21}$ \\ \hline
        $\frac{\Delta m_{3\ell}^{2}}{10^{-3}eV^{2}}$ & $\left(A-\frac{B}{A}\right)\approx 2.523_{-0.030}^{0.032}$ & $\frac{A}{2}\approx 2.509_{-0.030}^{+0.032}$ \\ \hline
\end{tabular}
\caption{Conditions for reproducing the neutrino squared mass differences for normal and inverse ordering.}
\label{massdifderences}
\end{table}
  
The discrete symmetry forces the lightest neutrino to be tree-level massless. It is the particular $\mathbb{Z}_{2}$ parity of these inert $\sigma$ and $\sigma'$  superfields that allows a one-loop finite correction. Furthermore, just like the electron acquire a finite mass via radiative corrections, the lightest neutrino might acquire a mass as well. However, they are not taken into account since we are interested in the squared mass differences and we can consider radiative corrections to be of the same order so they cancel out in the differences. 
  
\section{Family mixing}

In the previous section, lepton mass spectrum was determined analytically considering the observed physical masses hierarchy. Nevertheless, Yukawa couplings must also ensure the PMNS matrix. There have been some works on fermion mass structures in order to determine the cases in which masses can be reproduced \cite{neutrinoTex}, \cite{MTex} as well as CKM and PMNS matrices \cite{CKMcan}, \cite{PMNScan}. In the present work, the problem of massless tree-level particles is overcome thanks to radiative corrections induced by the $\sigma$ and $\sigma'$ scalars and exotic fermions which must also be considered for an appropriate reproduction. 

Now, we determine the number of free parameters in the charged lepton rotation relevant to the PMNS matrix reproduction. On the one hand, the $3\times 3$ submatrix concerning SM particles the charged lepton rotation in Eq. (\ref{VL}) depends on $\theta_{e\mu}$, $\theta_{e\tau}$, $\tilde{\Sigma}$, $v_{2}$, $v_{2}^{\prime}$, $h_{\sigma e}^{E e}$ and $h_{\sigma e}^{E \tau}$ but the parameter space can be reduced. First, we fix the VEV according to lepton masses as explained in \cite{model}, being $v_{2}=138$ GeV and $v'_{2}=21$ GeV. Then, the electron mass scale is set by $\tilde{\Sigma}=10^{-6}$ while $h_{\sigma e}^{E e}$ and $h_{\sigma e}^{E \tau}$ are restricted by the electron mass. In general, either $h_{\sigma e}^{E e}$ or $h_{\sigma e}^{E \tau}$ can be taken as a free parameter but they enter in the rotation matrix through $\Omega_{2}$ and gets multiplied by $\tilde{\Sigma}$, allowing us to neglect its influence on the rotation matrix. Finally, the $\theta_{e\tau}$ dependence is removed by considering the restriction of the Eq. (\ref{emuetau}) so we consider $\theta_{e\mu}$ as the single relevant free parameter from the charged lepton rotation.

On the other hand, neutrino rotation matrix can be obtained from the PMNS matrix and charged lepton rotation so the relevant free parameters are the $6$ couplings $h_{2e(\mu)}^{\nu k}$ $k=e,\mu,\tau$, present in the active neutrino mass matrix. Such parameters are obtained by doing a numerical exploration compatible with lepton masses, neutrino squared mass differences and the PMNS matrix, for a given $\theta_{e\mu}$.  It is initially found that the minimum number of complex parameters in the neutrino mass matrix is three. Therefore, we choose as complex parameters the three $\tilde{h}_{2e}^{\nu k}$ couplings, $k=e,\mu,\tau$, whose complex phase is introduced by the following replacement: 

\begin{align}
    \tilde{h}_{2e}^{\nu e} &\rightarrow  \tilde{h}_{2e}^{\nu e}e^{i\alpha}, & \tilde{h}_{2e}^{\nu \mu} &\rightarrow  \tilde{h}_{2e}^{\nu \mu}e^{i\beta}, & \tilde{h}_{2e}^{\nu \tau} &\rightarrow  \tilde{h}_{2e}^{\nu \tau}e^{i\gamma}, \nonumber \\
    \tilde{h}_{2\mu}^{\nu e} &\rightarrow  \tilde{h}_{2\mu}^{\nu e}, & \tilde{h}_{2\mu}^{\nu \mu} &\rightarrow  \tilde{h}_{2\mu}^{\nu \mu}, & \tilde{h}_{2\mu}^{\nu \tau} &\rightarrow  \tilde{h}_{2\mu}^{\nu \tau}, \label{choice}  
\end{align}
\noindent
where $\tilde{h}_{2e(\mu)}^{\nu k}$ are taken as real numbers and the complex phases are parametrized less than $\pi/2$. Such a choice is equivalent to making all $h_{2\mu}^{\nu k}$ parameters as complex since the active neutrino mass matrix in Eq. (\ref{mlight}) is symmetric under the replacement $h_{2e}^{\nu k} \leftrightarrow h_{2\mu}^{\nu k}$.  Thus, the numerical fitting of the $\tilde{h}_{2e}^{\nu k}$ and $\tilde{h}_{2\mu}^{\nu k }$ parameters, as chosen according to Eq. (\ref{choice}), is done in a Mathematica script that looks for solutions to the equation:
 
 \begin{align}
      m_{light} =& V^{\ell T}_{L-SM}U^{*}m_{light}^{diag}U^{\dagger}V^{\ell}_{L-SM}, \label{mlightnum}
 \end{align}
 
 \noindent
where $V^{\ell}_{L-SM}$ is the $3\times 3$ submatrix of the charged lepton rotation $V_{L}^{\ell}$ in Eq. (\ref{VL}), containing the mixing of the SM particles:

\begin{align}
    V^{\ell}_{L-SM}&=\begin{pmatrix}
       \cos\theta_{e\mu} & \sin\theta_{e\mu} & - \frac{v_{2}}{v'_{2}}\Omega_{2}\tilde{\Sigma}\cos\theta_{e\mu} \\
 -\sin\theta_{e\mu} & \cos\theta_{e\mu} & \frac{v_{2}}{v'_{2}}\Omega_{2}\tilde{\Sigma}\sin\theta_{e\mu} \\
 \frac{v_{2}}{v'_{2}}\Omega_{2}\tilde{\Sigma} & 0 & 1 
    \end{pmatrix}.
\end{align}
\noindent
Moreover, $m_{light}$ simplifies to:
\small
 \begin{align} \label{mlight}
    m_{light} \equiv 
    &\left( 
\begin{matrix}
	(\tilde{h}_{2e}^{\nu e})^{2}e^{2i\alpha} + (\tilde{h}_{2\mu}^{\nu e})^{2}  &
	{\tilde{h}_{2e}^{\nu e}}\,{\tilde{h}_{2e}^{\nu \mu}}e^{i(\alpha + \beta)} + {\tilde{h}_{2\mu}^{\nu e}}\,{\tilde{h}_{2\mu}^{\nu \mu}} 	&
	{\tilde{h}_{2e}^{\nu e}}\,{\tilde{h}_{2e}^{\nu \tau}}e^{i(\alpha + \gamma)}+ {\tilde{h}_{2\mu}^{\nu e}}\,{\tilde{h}_{2\mu}^{\nu \tau}} 	\\
	{\tilde{h}_{2e}^{\nu e}}\,{\tilde{h}_{2e}^{\nu \mu}}e^{i(\alpha + \beta)} + {\tilde{h}_{2\mu}^{\nu e}}\,{\tilde{h}_{2\mu}^{\nu \mu}}	&	
	( \tilde{h}_{2e}^{\nu \mu})^{2}e^{2i\beta} +  (\tilde{h}_{2\mu}^{\nu\mu})^{2} 	&	
	{\tilde{h}_{2e}^{\nu \mu}}\,{\tilde{h}_{2e}^{\nu \tau}}e^{i(\beta + \gamma)}+ {\tilde{h}_{2\mu}^{\nu \mu}}\,{\tilde{h}_{2\mu}^{\nu \tau}}	\\
	{\tilde{h}_{2e}^{\nu e}}\,{\tilde{h}_{2e}^{\nu \tau}}e^{i(\alpha + \gamma)}+ {\tilde{h}_{2\mu}^{\nu e}}\,{\tilde{h}_{2\mu}^{\nu \tau}}	&	
	{\tilde{h}_{2e}^{\nu \mu}}\,{\tilde{h}_{2e}^{\nu \tau}}e^{i(\beta + \gamma)}+ {\tilde{h}_{2\mu}^{\nu \mu}}\,{\tilde{h}_{2\mu}^{\nu \tau}}	&	
	 (\tilde{h}_{2e}^{\nu \tau} )^{2}e^{2i\gamma} +  (\tilde{h}_{2\mu}^{\nu \tau})^{2} 
\end{matrix} \right), 
 \end{align}
\normalsize
\noindent
 and $m_{light}^{diag}$ is the mass matrix in the diagonal basis, which reads $m_{light}^{diag}=diag(0,m_{2},m_{3})$ for normal ordering and $m_{light}^{diag}=diag(m_{1},m_{2},0)$ is for inverse ordering. It is worth to remember that such diagonal mass matrix comes as a consequence of having a zero mass eigenstate contained in the mass matrix of Eq. (\ref{mlight}), which leads to definite masses $m_{1}$, $m_{2}$ and $m_{3}$ as shown in table \ref{diagonalnumasses}.

\begin{table}[H]
    \centering
    \begin{tabular}{|c|c|c|c|} \hline
           &  $m_{1}$ & $m_{2}$ & $m_{3}$ \\ \hline
        NO & $0$ & $\sqrt{m_{21}^{2}}$ & $\sqrt{m_{3l}^{2}}$  \\ \hline
        NO & $0$ & $\sqrt{m_{21}^{2}}$ & $\sqrt{m_{3l}^{2}+m_{21}^{2}}$ \\ \hline
        IO & $\sqrt{-m_{3l}^{2}}$ & $\sqrt{m_{21}^{2}-m_{3l}^{2}}$ & $0$ \\ \hline
        IO & $\sqrt{-m_{3l}^{2}-m_{21}^{2}}$ & $\sqrt{-m_{3l}^{2}}$ & 0 \\ \hline
    \end{tabular}
    \caption{Neutrino mass eigenvalues for Normal and Inverse Ordering for a theory with one massless neutrino.}
    \label{diagonalnumasses}
\end{table}

\begin{figure}[H]
    \centering
    \includegraphics[scale=0.2]{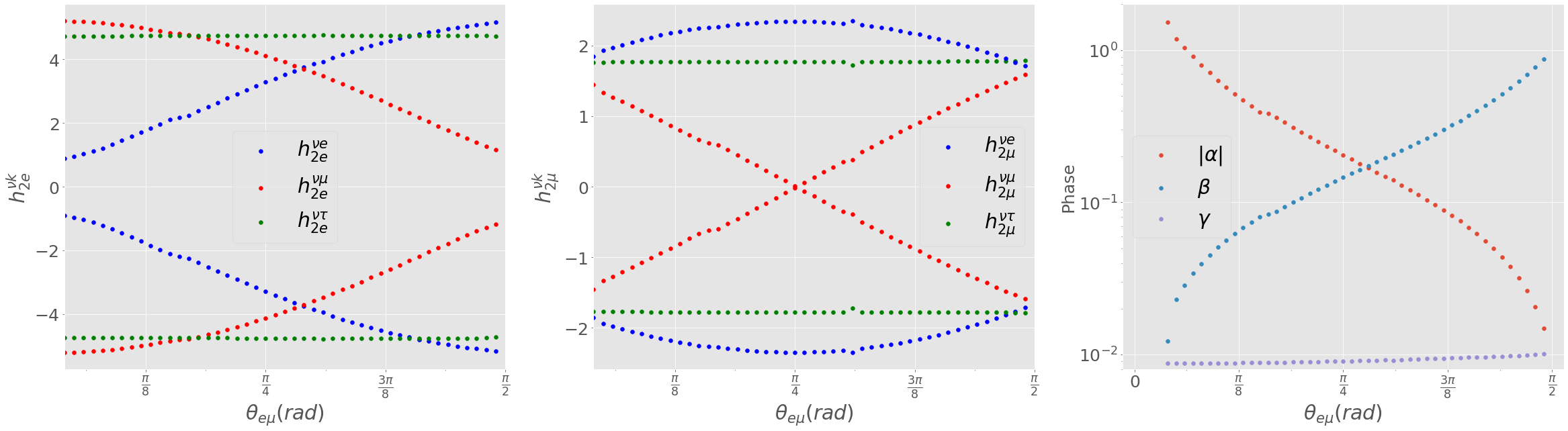}
    \caption{Neutrino Yukawa couplings and phases values as a function of $\theta_{e\mu}$ for the Normal Ordering Scheme.}
    \label{NOresults}
\end{figure}

\begin{figure}[H]
    \centering
    \includegraphics[scale=0.2]{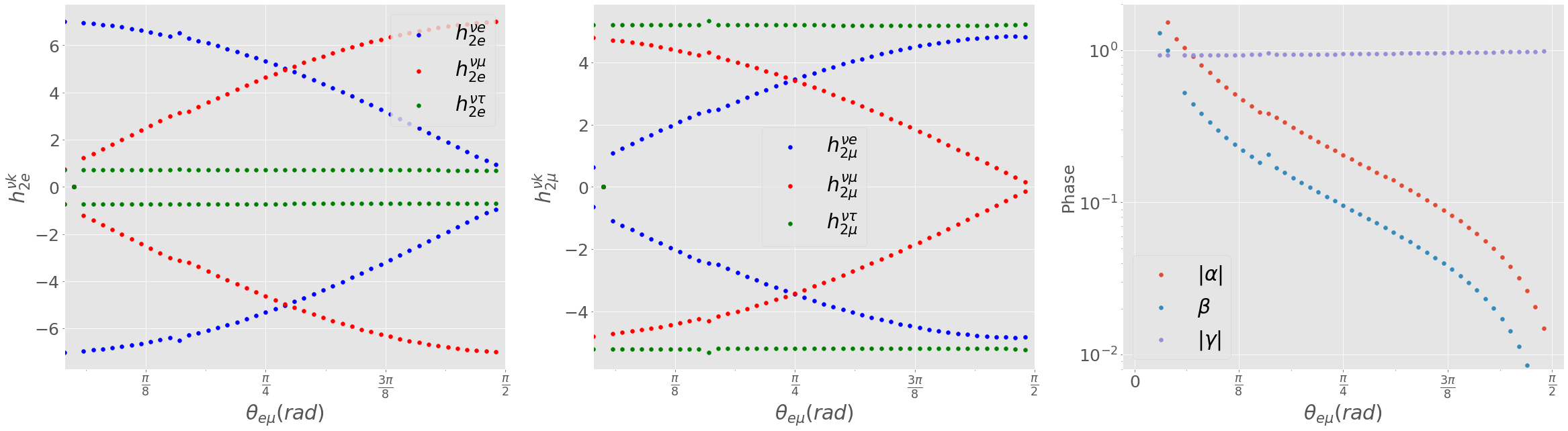}
    \caption{Neutrino Yukawa couplings and phases values as a function of $\theta_{e\mu}$ for the Inverse Ordering Scheme.}
    \label{IOresults}
\end{figure}

The results of the numerical fitting to the PMNS matrix gives as output the allowed values for $\tilde{h}_{2e(\mu)}^{\nu k}$ and the complex phases $\alpha, \beta, \gamma$. It is found that neutrino Yukawa couplings have a strong dependence on $\theta_{e\mu}$ for both Normal and Inverse Ordering, as shown in figures \ref{NOresults} and \ref{IOresults}. Besides, it was obtained that all $\tilde{h}_{2e(\mu)}^{\nu k}$ parameters are at the $ \text{meV}^{1/2}$ scale, such order of magnitude can be connected to $h_{2e(\mu)}^{\nu k}$ by an overall factor of order $\frac{\mu_{N} v_{2}^{2}}{{h_{N\chi 1}}^{2}v_{\chi}^{2}} \approx 1$ meV while Yukawa couplings $h_{2e(\mu)}^{\nu k}$ are taken of order 1.

We present in figures \ref{NOresults} and \ref{IOresults} the allowed values for $h_{2e(\mu)}^{\nu k}$ as a function of $\theta_{e\mu}$ for Normal and Inverse ordering respectively. It can be seen that there are two solutions for each Yukawa coupling, a positive and a negative, although the sign of each coupling cannot be chosen independently of the others. On the one hand, Normal Ordering scheme allows all Yukawa couplings to be either positive or negative while $\alpha$ phase is negative, so its absolute value is shown. On the other hand, Inverse Ordering scheme requires either $h_{2\mu}^{\nu \mu}$ negative and $h_{2\mu}^{\nu e}$ and $h_{2\mu}^{\nu \tau}$ with positive sign, or $h_{2\mu}^{\nu e}$ and $h_{2\mu}^{\nu \tau}$ to be negative while $h_{2\mu}^{\nu \mu}$ positive, forbidding the same sign for all three couplings although again both positive and negative value is shown in figure \ref{IOresults}. Furthermore, $\alpha$ and $\gamma$ phases are negative in this case, so their absolute value is shown. 

On the whole, we have obtained that given the neutrino mass eigenvalues there is always a set of dimensionless Yukawa couplings that recreate the PMNS matrix.  In this case, three complex parameters and three real parameters from neutral lepton interactions together with one mixing angle from charged lepton rotation, are the minimal set of parameters needed to reproduce the PMNS matrix with no additional Majorana phases.

\section{Conclusions}

Abelian extensions to the MSSM have been widely studied and often involve heavy particles. In this case, exotic fermions, additional scalars and non-universality build an anomaly free theory in which we can identify three energy scales: electroweak, $U(1)_{X}$ and SUSY breaking scale. In fact, electroweak scale provides the masses of SM leptons while $U(1)_{X}$ scale is related to exotic particles which are important to generate the radiative corrections to the electron mass and charged lepton rotation matrix, although Majorana neutrinos provide small active neutrino masses via inverse seesaw mechanism.

In the lepton sector, non-universality led to a particular lepton mass matrix texture where the electron is massless at tree-level, but acquires a finite mass value via inert scalar singlets, exotic fermions and its superpartners at one-loop level. Additionally, the mass matrix diagonalization led to $\mu$ and $\tau$ masses at tree-level depending exclusively on $v'_{2}$.

Likewise, non-universality led to one massless neutrino which allows to get a definite mass spectrum in both NO and IO schemes. The active neutrinos acquire masses at the meV scale via inverse-seesaw mechanism which also generates six Majorana neutrinos with masses at the order of $v_{\chi}$. 

Lastly, from a parameter fit to the PMNS matrix it is found that the rotation matrix of left-handed charged leptons and neutral lepton Yukawa couplings $h_{2e(\mu)}^{\nu k}$ can be parametrized by $\theta_{e\mu}$, which is the only relevant free parameter in the model since $\theta_{e\tau}$ was related to $\theta_{e\mu}$ by the Eq. (\ref{emuetau}). Besides, there is a sign restriction of such couplings whether Normal or Inverse Ordering is considered and the $h_{2e(\mu)}^{\nu k}$ allowed values shows that $\tau$-related couplings tend to have a fixed value, both in magnitude and complex phase.

\section{References}
\bibliographystyle{spphys}

\end{document}